\documentclass[useAMS,usenatbib,psfig,eps,graphics]{mn2e} 
\usepackage{epsfig, graphicx,natbib} 
 
\def\grtsim{\mathrel{\hbox{\rlap{\hbox{\lower2pt\hbox{$\sim$}}}\raise2pt\hbox{$>$}}}} 
\def\lesssim{\mathrel{\hbox{\rlap{\hbox{\lower2pt\hbox{$\sim$}}}\raise2pt\hbox{$<$}}}}

\def\degree{\nobreak\ifmmode{^\circ}\else{$^\circ$}\fi}

\newcommand{\aap}{A\&A}

\newcommand{\apj}{ApJ} 
 
\newcommand{\apjs}{ApJS} 
 
\newcommand{\araa}{ARA\&A} 
\newcommand{\mnras}{MNRAS} 
\newcommand{\nat}{Nat}

\begin{document} 
\topmargin -0.5in %this is only for astro-ph, uncomment when submitting paper 
 
\title[Evidence for  high-z Compton-thick quasars]{Evidence for a large
  fraction of Compton-thick quasars at high redshift } \author[A. Mart\'\i nez-Sansigre et al. ]{Alejo Mart\'\i nez-Sansigre$^{1,2}$\thanks{E-mail: 
martinez@mpia-hd.mpg.de (AMS)}, Steve Rawlings$^{1}$, David G. Bonfield$^{1}$,  Silvia Mateos$^{3}$, \and  Chris Simpson$^{4}$, Mike Watson$^{3}$, Omar Almaini$^{5}$, S\'ebastien Foucaud$^{5}$,\and Kazuhiro Sekiguchi$^{6}$,  Yoshihiro Ueda$^{7}$ \\  
\footnotesize\\  
$^{1}$Astrophysics, Department of Physics,University of Oxford, Keble Road, Oxford OX1 3RH, UK\\ 
$^{2}$Max-Planck-Institut f\"ur Astronomie, K\"onigstuhl-17, D-69117,
Heidelberg, Germany \\
$^{3}$Department of Physics and Astronomy, University of Leicester, Leicester, LE1 7RH, UK\\ 
$^{4}$Astrophysics Research Institute, Liverpool John Moores University, Twelve Quays House, Egerton Wharf, Birkenhead CH41 1LD, UK\\ 
$^{5}$School of Physics and Astronomy, University of Nottingham, University Park, Nottingham NG7 2RD, UK\\ 
$^{6}$Subaru Telescope, National Astronomical Observatory of Japan, 650 North A'ohoku Place, Hilo, Hawai'i 96720, USA\\ 
$^{7}$Department of Astronomy, Kyoto University, Kyoto 606-8502, Japan\\ 
}

\date{} 
 
\pagerange{\pageref{firstpage}--\pageref{lastpage}} \pubyear{} 
 
\maketitle 
 
\label{firstpage} 
\vspace{-0.5 cm} 
 
\begin{abstract} 
  Using mid-infrared and radio selection criteria, we pre-select a sample of
  candidate high-redshift type-2 quasars in the Subaru \textit{XMM-Newton\/} Deep Field
  (SXDF).  To filter out starburst contaminants, we use a bayesian method to
  fit the spectral energy distributions (SEDs) between 24-$\mu$m and B-band,
  obtain photometric redshifts, and identify the best candidates for high-$z$
  type-2 quasars.  This leaves us with 12 $z_{\rm phot} \geq 1.7$ type-2
  quasar candidates in an area $\sim 0.8$ deg$^{2}$, of which only two have
  secure X-ray detections. The two detected sources have estimated column
  densities $N_{\rm H}\sim$2 \& $3\times10^{27}$ m$^{-2}$, i.e. heavily
  obscured but Compton-thin quasars. Given the large bolometric luminosities
  and redshifts of the undetected objects, the lack of X-ray detections
  suggests extreme absorbing columns $N_{\rm H} \grtsim 10^{28}$ m$^{-2}$ are
  typical.  We have found evidence for a population of ``Compton-thick''
  high-redshift type-2 quasars, at least comparable to, and probably larger
  than the type-1 quasar population, although spectroscopic confirmation of
  their AGN nature is important.
\end{abstract} 
 
\begin{keywords} 
quasars:general-galaxies:nuclei-galaxies:active-X-rays:galaxies 
\end{keywords} 
 
\section{Introduction}

High-redshift quasars with column densities $N_{\rm H}\grtsim10^{28}$ 
m$^{-2}$ are so heavily absorbed that they are barely detectable in even 
 the most sensitive hard X-ray surveys \citep[see, 
e.g.,][]{2005ARA&A..43..827B}, and are known as 
``Compton-thick''. While in the local Universe, 40\% of active 
galactic nuclei (AGN) are found to be Compton-thick \citep 
{1999ApJ...522..157R}, the fraction of high-redshift quasars with such 
absorbing columns is currently unknown.  
 
An alternative to (hard) X-ray selection is in the mid-infrared, where the
obscuration due to dust becomes small. AGN are powerful mid-infrared emitters
due to dust surrounding the accretion disk (the torus) reprocessing the UV and
X-ray photons from the central engine.  AGN invisible in X-rays were indeed
found by \citet {2005ApJ...634..169D}, who used a mid-infrared and radio
excess criterion to to select a sample of AGN. The majority of these AGN are
at $z \lesssim 1$ and are better described as Seyfert-2s by virtue of their
low luminosities. The authors found
that while the sources detected in the X-ray were unlikely to be
Compton-thick, the AGN not detected in the X-ray ($\sim$20\% of their sample)
could be Compton-thick ($N_{\rm H}$$\grtsim10^{28}$ m$^{-2}$). The obscured
and gravitationally-lensed quasar IRAS FSC\,10214+4724, at $z\sim2.3$ also
appears to be Compton-thick \citep {2005MNRAS.357L..16A}.
 
\citet {2006ApJ...642..673P} use two different selection criteria,
X-ray and infrared, to look for Compton-thick AGN in an area of 0.6
deg$^{2}$. Their X-ray selection, with a flux limit $\sim$10$^{-17}$ W
m$^{-2}$ in the 2.5-8 keV band, finds 5 Compton-thick AGN. Of these 2
are spectroscopically confirmed high-redshift quasars.  Of the
infrared-selected sources, the strict SED criteria imposed on the
sources guarantee AGN, but are likely to exclude heavily obscured
($A_{\rm V}\grtsim 20$) high-redshift ($z\geq1$) sources which do not
show power-law mid-infrared spectral energy distributions (SEDs), but
have significant contributions from stellar light at 3.6 and 4.5
$\mu$m.  It is not suprising, therefore, that they find a smaller
Compton-thick fraction ($\sim$10\%) than expected in their
infrared-selected sample, as it is probably biased against heavily
obscured AGN. At high luminosities and redshift, Alonso-Herrero et
al. (2006, AH06) find $\sim$50\% of their sources are undetected in
X-rays (although not necessarily Compton-thick). This sample is based
on a power-law criterion, and could plausibly also be missing the most
heavily obscured sources.

\citet {2005Natur.436..666M}, hereafter MS05, found a population of
high-redshift type-2 quasars at least as numerous as the unobscured (type-1)
population and which possibly outnumber the type-1s by $\sim$2-3:1. The most
recent X-ray studies agree with a large obscured fraction but disagree on the
relative numbers of Compton-thick sources: \citet {2006MNRAS.372.1755D} find a 
$\sim$3:1 ratio amongst Compton-thin type-2 and type-1 quasars, while Gilli,
Comastri \& Hasinger (2006) infer a $\sim$1:1 ratio of Compton-thin type-2 to
type-1 quasars, although the total type-2 to type-1 ratio can increase to 
$\sim$2:1 when Compton-thick sources are included.

In this letter, we present the X-ray properties of a sample of obscured
(type-2) high-redshift quasars, selected from mid-infrared and radio data in
the Subaru \textit{XMM-Newton\/} Deep Field (SXDF).

\section{Sample selection and dataset}

We proceed to select a similar sample to that of MS05, in the SXDF, where deep
X-ray data are available. In this work we decrease the lower radio flux
density criterion, to increase the number of candidates. This is important as
here we have a smaller area ($\sim$0.8 deg$^{2}$ as opposed to 3.8 deg$^{2}$
in the MS05 sample). The selection criteria used here are $S_{24~\mu \rm m} >
300~\mu$Jy, $S_{3.6~\mu \rm m} \leq 45~\mu$Jy, and 100 $\mu$Jy $\leq S_{1.4
  \rm GHz} \leq$ 2 mJy, which yields 38 candidates. For a detailed discussion
of these criteria see \citet{2006MNRAS.tmp..691M}.

As a brief summary, the two mid-infrared criteria are able to target 
$z\sim2$ type-2 quasars with $A_{\rm V}\grtsim 5$, but will also allow 
$z\lesssim 1$ ultra-luminous infrared galaxies (ULIRGs) in the sample. 
These ULIRGs will have radio luminosities following the FIR-radio 
correlation \citep{1992ARA&A..30..575C}, and so by choosing a 
high-enough radio flux, one can cut out all but the most extreme 
starburst contaminants.  The MS05 criteria were carefully chosen to 
avoid starburst galaxies, and lowering the radio criterion has the 
disadvantage that less extreme starbursts are allowed in the sample.

The mid-infrared data were obtained by using the SWIRE DR2
\citep{2005AAS...207.6301S}, which covers the SXDF, and has a flux density
limit of $\sim$250 $\mu$Jy at 24 $\mu$m (5$\sigma$), and 10 $\mu$Jy for both
3.6 and 4.5 $\mu$m (10$\sigma$ and 5$\sigma$ respectively).  The 1.4 GHz
catalogue used is described in \citet {Simpson:2006if} and comes from a
B-array ($\sim$5$\times$4 arcsec$^{2}$ beam) VLA survey with a peak flux
density limit of 100 $\mu$Jy ($\sim$5-8 $\sigma$).

The spectroscopic completeness of the SXDF are poor at this stage, so we
decided to undertake this preliminary study using photometric redshifts (see Section~3). %This
%was much helped by the fact that, for most of our target objects, the optical
%to near-infrared light originates from the host galaxy only, as the quasar is
%severely obscured at these wavelengths (except possibly the object ID135, see
%Table~1 and supplementary Figures
%\footnote{http://www.mpia.de/homes/martinez/sxdftype2seds.pdf}). 
%The dataset
%for this came from two sources. 
The optical data were obtained from the Subaru
B,V,R,i' and z' imaging (Furusawa et al., in prep.). The near-infrared data (J
and K) were taken from the UKIDSS ultra-deep survey (UDS) DR1 \citep {Warren:2006uz}. The
X-ray data were obtained with the \textit{XMM-Newton\/} observatory, which
covered the SXDF field in 7 pointings, with exposures of 50 ks each, except
for the central pointing, which was observed for 100 ks (Ueda et al., in
prep.).

\begin{table*} 
\begin{center} 
\begin{tabular}{llllllrrrrr} 
\hline 
\hline 
Name &  RA  & Dec & $z_{\rm phot}$  & log$_{10}$[$L_{\rm bol}$ & $A_{\rm V}$     & log$_{10}$[$L_{\rm gal}$   &log$_{10}$[$L_{\rm blue}$ & ln[OR(q/u)] & log$_{10}$[p(d$|$q)/ & $S_{2-12 \rm keV}$         \\ 
  & (J2000)   &  &  & /W] &  & /$L^{*}_{\rm k}$]  & /$L^{*}_{\rm k}$]  & & ($S_{17}$)$^{12}$/]    &/ $S_{17}$     \\ 
\hline 
ID052  & 02 16 17.92 & -05 07 18.56 &  1.90$\pm0.05$ & 41.2$^{1}$  & 119 & 0.85 & 0.55 &13.0&-10.0 & $<$0.7    \\ 
ID123 & 02 19 28.76 & -05 09  08.81 &  1.75$\pm0.05$ & 40.2  & 72 & 0.65 & 0.20 & 5.2 & -9.7 & $<$2.9$^{2}$    \\ 
ID135 & 02 19  01.89 & -05 11 14.22 &  4.15$_{-0.09}^{+0.05}$ & 40.2  & 2.2$^{3}$ & 1.20 & 0.15 &26.0& -9.4& $<$0.3    \\ 
ID142 & 02 17 23.82 & -04 35 13.72 &  4.05$\pm0.05$ & 40.4  &  6.7 & 1.25 & 0.45 &10.0& -8.9& $<$0.4    \\ 
ID147 & 02 19 10.31 & -05 16  03.00 &  1.80$\pm0.05$ & 40.1  & 35.9 & 0.75 & -0.05 &28.6& -13.4& $<$0.3    \\ 
ID200 & 02 18 15.71 & -05 05 10.34 &  1.75$_{-0.07}^{+0.05}$ & 40.1  & 72 & 0.75 & 0.50 & 8.4& -9.1& $<$0.4    \\ 
ID249 & 02 19 13.74 & -04 56 04.27   &  1.75$\pm0.05$ & 40.0 & 35.9 & 0.50 & 0.00 & 25.2& -9.2 & $<$0.6 \\ 
ID342 & 02 17  05.35 & -05 09 24.61 &  3.85$\pm0.05$ & 40.0 & 9.0 & 1.20 & 0.30 & 4.6 & -15.25 & $<$6.8$^{2}$ \\
ID345 & 02 16 29.56 & -05 03 10.65 &  2.00$_{-0.07}^{+0.08}$ & 39.8  & 35.9$^{4}$ & 1.00 & 0.10 & 10.4& -11.1&  $<$0.6  \\ 
ID347 & 02 18  09.64 & -05 18 42.42 &  1.75$\pm0.05$ & 40.2  & 72  & 0.55 & -0.30 &5.8& -8.0 & 1.7$\pm$0.9$^{5}$    \\ 
ID386 & 02 17 25.11 & -05 16 17.27 &  1.90$_{-0.12}^{+0.05}$ & 39.4  & 17.9 & 0.50 & -0.75 &6.2& -11.5& $<$0.1    \\ 
ID401 & 02 16 23.02 & -05 08  06.76 &  1.90$_{-0.12}^{+0.05}$ & 40.0   & 35.9 & 0.70 & 0.40 &26.7& -8.9 &  3.4$\pm$1.3$^{5}$    \\ 
\hline
\hline 
\end{tabular} 
\caption{ \noindent Best-fit parameters and X-ray fluxes for the high-redshift
  type-2 quasars. The names are from the radio catalogue of \citet
  {Simpson:2006if}, and we define $S_{17} = 10^{-17}$ W m$^{-2}$ for
  convenience.  The errors in $z_{\rm phot}$ are estimated, given our model
  SEDs, from the full-width half-maximum values from the marginalised
  probability distribution functions (PDFs) for $z_{\rm phot}$. For most
  objects, the PDF has at least one secondary peak, always adjacent to the
  primary one, with peak value between 0.5 and 0.05 of the primary peak.  The
  median value of log$_{10}$[$L_{\rm gal}$/$L^{*}_{\rm k}$], 0.75, corresponds
  to a 5.6$L^{*}_{\rm k}$ (where $L^{*}_{\rm k}$ is the break in the local
  K-band luminosity function) galaxy at $z\sim2$, which asuming passive
  evolution and using the models of \citet{2003MNRAS.344.1000B}, would become
  a present-day $\sim$$L^{*}_{\rm k}$ galaxy. The limits quoted for X-ray
  non-detections are the flux returned by the task EMLDETECT at the position
  of the source, plus the error quoted for this flux. This is approximately
  the same as a 2$\sigma$ limit.  $^{1}$This high value for $L_{\rm bol}$ is
  probably due to an overestimation of the $A_{\rm V}$.  $^{2}$The reason for
  these high values for the fluxes is that the background noise is
  significantly higher in these two sources.  $^{3}$This value of $A_{\rm V}$
  suggests a reddened type-1 quasar rather than a genuine type-2.  $^{4}$ID345
  is a point source at K-band, which suggests the value of $A_{\rm V}$ is an
  overestimate.  $^{5}$ID347 and ID401 have safe X-ray detections.  The
  best-fitting SEDs for all 12 objects can be found at:
  http://www.mpia.de/homes/martinez/sxdftype2seds.pdf}
\label{tab:xtable} 
\end{center} 
\end{table*}

\section{Photometric redshifts and filtering out ULIRG contaminants}
 
In order to disentangle type-2 quasars from ULIRGs and obtain 
photometric redshifts, the SEDs of the candidates were fitted using a 
model consisting of three components (warm dust, galaxy light and blue 
light). The normalisations of the components were allowed to vary 
together with the redshift, $z$, although for a given $A_{\rm V}$ and 
$z$, the normalisation of the warm dust component was fixed to fit the 
24 $\mu$m data. Two different models were investigated, and the bayesian odds ratio \citep[e.g.][]{Sivia:96} was used to select 
between the models. These models differed only in their warm dust 
properties and consisted of a quasar (with dust) and a ULIRG.  The 
redshift was given a flat prior between $0 \leq z \leq 7$. No real 
quasars have been detected at $z \grtsim 7$, justifying this cutoff. 
 
The stellar population was modelled by taking the SED of a $z=0$ 
elliptical galaxy from Coleman, Wu \&    Weedman (1980).  SEDs showing 
smaller 4000-\AA~breaks were well fitted by the elliptical galaxy 
together with the blue component (see below). The elliptical was 
normalised to match the K-band luminosity of an $L^{*}$ galaxy in the 
local K-band luminosity function of \citet {2001MNRAS.326..255C}. The 
luminosity of the galaxy relative to the local $L^{*}_{\rm K}$ was 
then allowed to vary with a prior flat in log-space, between $-1.3 
\leq$log$_{10}(L_{\rm gal}/L^{*}_{\rm K})$$\leq 1.3$. This choice 
brackets the reasonable range of host galaxies for a powerful quasar 
accounting for passive evolution: from a very faint $0.05L^{*}_{\rm 
K}$ to a very bright $20L^{*}_{\rm K}$ galaxy. The choice of a prior 
flat in log-space means that the host galaxy is believed, a priori, to 
be as likely to lie between $0.1L^{*}_{\rm K}$ and $L^{*}_{\rm K}$ as 
between $L^{*}_{\rm K}$ and $10L^{*}_{\rm K}$. This is more realistic 
than a prior flat in real space, which would imply a quasar is as 
likely to be hosted by an $L^{*}_{\rm K}$ galaxy as by a $10L^{*}_{\rm 
K}$ one. 
 
The blue component, between 912~\AA\, and 5000~\AA, has a physical 
motivation as well as serving a practical purpose. It can represent 
the scattered light from the obscured quasar, blue light from young 
stars or a UV-upturn brighter than that of the template elliptical 
galaxy used (and therefore a smaller 4000-\AA~break). The slope was 
therefore set to $S_{\nu}\propto \nu^{-0.5}$, which is representative 
of both type-1 quasars and starforming galaxies. The normalisation of 
the blue component is given as a ratio of the luminosity to that of an 
$L^{*}_{\rm K}$ galaxy at 5000~\AA, and was allowed to vary 
independently of the galaxy, with a flat prior between $-2 
\leq$log$_{10}(L_{\rm blue}/L^{*}_{\rm K,5000 \AA})$$\leq 2$, spanning 
the range between a blue component fainter than an $L^{*}_{\rm K}$ 
elliptical galaxy's UV-upturn, and as bright as the most powerful 
type-1 quasar. To avoid overfitting and losing the ability to discriminate
between models, neither the stellar or the blue component are reddened by dust.

The ULIRG dust component was modelled using the models of \citet 
{Siebenmorgen:2007ca}, hereafter SK07, only allowing variation of one 
parameter, $A_{\rm V}$. For a given  $A_{\rm V}$ and $z$, the bolometric
luminosity of the ULIRG, $L_{\rm bol}$ was chosen to make the SED go as close
as possible to the 24-$\mu$m data point. The values of $L_{\rm bol}$ are only 
restricted by the range available in the SK07 library: $10.1 \leq$ 
log$_{10}(L_{\rm bol}/L_{\odot})$ $\leq 12.7$ for our choice of 
parameters. Of the other parameters, the nuclear radius was fixed to 1 
kpc, the ratio of luminosity of OB stars with hot spots to $L_{\rm 
tot}$ was fixed to 0.6, and the hydrogen number density was fixed to 
$10^{10}$ m$^{-3}$. The models have a discreet set of values of the 
extinction ($A_{\rm V}=$ 2.2, 4.5, 6.7,9, 17.9, 35.9, 72 and 119), so 
the prior for the $A_{\rm V}$ consists of a set of $\delta$-functions 
(with equal probability) at these values.

The mid-infrared SED for quasars was modelled using the \citet 
{1994ApJS...95....1E}, hereafter E94, type-1 SED, and obscuring it 
with dust from the models of \citet{1992ApJ...395..130P}. For 
consistency with SK07, only Milky Way (MW)-type dust is used. This SED 
and dust model would allow us, in principle, to vary the values of 
$A_{\rm V}$ continuously, but to make the fitting procedure as fair as 
possible, we restricted the values of $A_{\rm V}$ to those present in 
the models of SK07 and assigned the same prior to them. This flat 
prior reflects our ignorance about the range in $A_{\rm V}$, 
particularly for quasars.  For each object, the quasar 
bolometric luminosity $L_{\rm bol}$ was not allowed to vary freely, 
but was fixed for a given $A_{\rm V}$ and $z$ by the observed flux 
density at 24 $\mu$m. There is, however, no upper limit set on $L_{\rm 
bol}$. 
 
For a given model, the likelihood of a given combination of parameters 
is given by the n-dimensional probability density function 
 
\noindent \begin{eqnarray} 
 {\rm p(data|model~A, {\it z}, {\it A_{\rm V}}, {\it L_{\rm gal}}, {\it L_{\rm blue}})}=  \nonumber 
\end{eqnarray} 
 
\noindent \begin{equation} 
{1 \over (2\pi)^{n/2} \displaystyle\prod^{n}_{i}\sigma_{i}} e^{-\chi_{\rm A}^{2}/2},
\end{equation}

\noindent where  
 
\noindent \begin{equation} 
\chi_{\rm A}^{2} = \displaystyle\sum^{n}_{i} {\rm (model~A_{\it i}- data_{\it i})}^{2}/\sigma_{\it i}^{2},  
\end{equation} 
 
\noindent model A$_{i}$ is the flux predicted by model A, given the
parameters, over a given waveband, data$_{i}$ is the observed flux at that
band, $\sigma_{i}$ is the measurement error in that band and {\it n} is the
number of bands. In the cases where $\sigma_{i}$ was smaller than 10\% of the
flux density of the object, it was set to 10\%.  To treat non-detections at a
particular band, we followed the following method: when the observed galaxy
flux density and the model fell below the flux density limit
(5$\sigma_{band}$, where $\sigma_{band}$ is the rms noise in that band), the
band made no contribution to $\chi^{2}$. When the model lay above the limit,
the object was assigned a flux density $S_{i}$ and an error $\sigma_{i}$ both
equal to half of the flux density limit (so $S_{i}=$$\sigma_{i}=
2.5\sigma_{band}$).  An undetected source would therefore only be
1$\sigma_{i}$ away from zero flux density, and from the limit. This
prescription was appropriate for our sources, selected to be faint and
therefore close to the limits in most of the bands, to avoid excessively
penalising nondetections in one band.
 
To select between models, we follow \citet 
{Sivia:96} in calculating the odds ratio (OR), 
 
\begin{equation} 
{\rm OR(A/B)} = {{\rm p(d|A)~p(A)} \over {\rm p(d|B)~p(B)}}, 
\end{equation}

\noindent where, for brevity, we refer to model A as A, model B as B, and data
as d. We assume that models A and B are equally likely \emph{a priori}, ~${\rm
  p}(A)/{\rm p}(B) = 1$, implying ignorance in the fraction of quasars c.f.
starbursting galaxies.  Thus the evidence for model y (where y $\in
\{A,B\}$) is proportional to p(d$|$y), which is simply the likelihood (from eq
1) integrated over the parameter space spanned by the priors, i.e.:

\noindent\begin{eqnarray} 
\noindent {\rm p(d|y)} =\nonumber
% \,\,\,\,\,\,\,\,\,\,\,\,\,\,\,\,\,\,\,\,\,\,\,\,\,\,\,\,\,\,\,\,\,\,\,\,\,\,\,\,\,\,\,\,\,\,\,\,\,\,\,\,\,\,\,\,\,\,\,\,\,\,\,\,\,\,\,\,\,\,\,\,\,\,\,\,\,\,\,\,\,\,\,\,\,\,\,\,\,\,\,  \nonumber \\ 
%\int e^{-\chi_{\rm y}^{2}/2}{\rm p}(z){\rm p}(A_{\rm V}){\rm p}(L_{\rm gal}){\rm p}(L_{\rm blue})\it {\rm d}z{\rm d}A_{\rm V}{\rm d}L_{\rm 
%gal}{\rm d}L_{\rm blue},  
\end{eqnarray} 
 
\noindent \begin{equation} 
\int e^{-\chi_{\rm y}^{2}/2}{\rm p}(z){\rm p}(A_{\rm V}){\rm p}(L_{\rm gal}){\rm p}(L_{\rm blue})\it {\rm d}z{\rm d}A_{\rm V}{\rm d}L_{\rm 
gal}{\rm d}L_{\rm blue}, 
\end{equation} 
  
\noindent $\chi_{\rm y}^{2}$ is as defined in Equation~2,  p($x$) is the normalised prior probability distribution of
parameter $x$ and the integral is over the entire prior space. To make sure
that our X-ray analysis did not include any ULIRGs, only objects with
OR(quasar/ulirg)$\geq 100$ were considered as quasars. This is equivalent to
stating we are at least 99\% confident of the quasar model over the ULIRG model. Note this criterion is
very similar to the so-called Jeffreys' criterion \citep {Jeffreys:1961}. In
addition, to make sure the fits were acceptable, only sources with evidence
p(data$|$quasar)$/(S_{17})^{12} \geq 10^{-17}$ were accepted, where $S_{17} =
10^{-17}$ W m$^{-2}$ ($S_{17}$ is defined to avoid numbers of order
$10^{-204}$ or smaller). Our range of evidences in Table~1,
$1.3\times10^{-9}\leq$p(data$|$quasar)$/(S_{17})^{12}$$\leq5.5\times10^{-16}$,
corresponds approximately to values of reduced $\chi^{2}=1.9-6.1$. Finally,
only sources with $z_{\rm phot} \geq 1.70$ were kept, since Ly~$\alpha$ is
visible in optical spectroscopy, so the redshifts can be checked
observationally, as long as this line is not obscured by dust on large scales
(see MS06). Preliminary spectroscopy of the sample suggests our values of
$z_{\rm phot}$ are in good agreement with spectroscopic redshifts (A. Mart\'\i
nez-Sansigre et al. in prep.).%, except for the objects with $z_{\rm phot} \sim
%4$, which are likely to lie at $z\sim3$.% with $z_{\rm phot}$ an overestimate
%due to the lack of H-band data.
 
\begin{figure}%[!h] 
\begin{center} 
\psfig{file=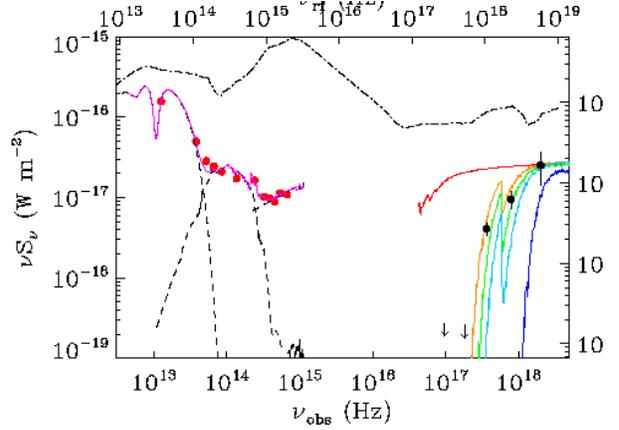,width=8cm,angle=0}  
\caption{\noindent Best fit spectral energy distribution (SED) of ID401. The
  top x-axis and right y-axis show the rest-frame frequencies and luminosities
  (in units of $L_{\odot}$) assuming the best-fit $z_{\rm phot}=1.90$. The
  dashed black lines show (from left to right) the quasar (with extinction),
  galaxy and blue templates normalised by the values (from
  Table~\ref{tab:xtable}) alongside the best fit to the data (red dots), the
  solid purple line shows the combined best-fit templates. The black dots and
  limits are the X-ray data in the 0.3-0.5, 0.5-1, 1-2, 2-4.5 and 4.5-12 keV
  bands, from all three EPIC cameras. The coloured lines are the WF99 absorbed
  models, with values of log$_{10}$($N_{\rm H}$/ m$^{-2}$) = 25.00 (red),
  27.25 (yellow), 27.50 (green), 27.75 (cyan) and 28.50 (dark blue), with
  normalisation so the unobscured curve (25.00) goes through the highest
  energy X-ray point. The X-ray datapoints are best described by the curve
  with 27.25 (yellow). For comparison, the E94 type-1 SED has been overplotted
  (dash-dotted line), with the same values of $L_{\rm bol}$ and $z_{\rm phot}$
  as ID401 (from Table~\ref{tab:xtable}). The low value of the unobscured WF99
  curve, compared to the E94 curve, shows this source has an intrinsically
  lower X-ray to mid-infrared ratio by a factor $\sim3$, than the expectation
  of E94.  }
\label{fig:sedID401} 
\end{center} 
\end{figure}

Of the 38 candidates in the sample, 12 (32\%) follow our above criteria and
are considered our best candidates for high-redshift type-2 quasars. Their
best-fit parameters are quoted in Table~\ref{tab:xtable}, and we proceed to
analyse the X-ray properties of these sources. Visual inspection of the
optical images showed several cases where the blue light was unresolved. This
fits in well with scattered light from the quasar reaching us and justifies
our choice of a ``blue'' component. Typically the best fitting normalisation
for the blue component is $\sim$5\% of the normalisation of the quasar
component, which is a reasonable scattering fraction. The K-band images
showed resolved sources in 11 cases, while ID345 is a point-source (suggesting
a real $A_{\rm V}\lesssim5$; a reddened type-1, see MS06). The SED and $A_{\rm
  V}$ of ID135 (see supplementary Figures) also suggest a reddened quasar,
rather than a genuine type-2, although since it is not point-like in K-band,
the host galaxy probably contributes significantly to the near-infrared flux.

\section{X-ray properties of the type-2 quasars}

The type-2 quasars were cross-matched with the X-ray catalogue of Ueda et al.
(in prep.) with a flux limit of $\sim$3$\times10^{-18}$ W m$^{-2}$ in the 2-12
keV band, but only 2 out of 12 (17\%) were detected: ID347 and ID401. These
objects have likelihoods (EP\_DET\_ML) of 81 and 80 respectively, and are
therefore clear detections (see DP06 for an explanation of the likelihood).

To obtain meaningful limits for the undetected sources, these had 
their counts measured directly from the X-ray image, using the 
\textit{XMM-Newton\/} Science Analysis Software (SAS). The positions of the 
sources not detected in the \textit{XMM-Newton\/} observations were added 
manually to the EBOXDETECT source lists of each observation. We then 
ran the SAS task EMLDETECT keeping the positions of the sources 
fixed. EMLDETECT performs a maximum likelihood fit on the distribution 
of observed counts of the sources previously extracted  by the task 
EBOXDETECT. The fit uses the five diferent energy bands (in keV: 
0.2-0.5,0.5-1,1-2,2-4.5,4.5-12) and the three EPIC cameras (M1,M2,pn) 
simultaneously. This procedure allowed to fit the sources as multiple 
components with separate point-spread functions (PSFs), and sources 
with no evidence of any X-ray detection were still fitted. 
 
This procedure allowed us to obtain fits to the counts for all sources in the
5 EPIC bands, and these were converted to fluxes in each band, assuming a
photon index $\Gamma=1.7$ and an obscuring column of $N_{\rm
  H}=3\times10^{24}$ m$^{-2}$ (the default SAS values). These are not the most
likely values for obscured quasars, so we estimate the effects of using more
representative values. Assuming
instead $\Gamma=1.9$, $N_{\rm H}=3\times10^{27}$ m$^{-2}$ at $z=2$ we estimate
the change in fluxes to be only +9\%, +3\% and -2\% in the 1-2, 2-4.5 and
4.5-12 keV bands, and therefore negligible compared to the large uncertainties
in the count rates. All the 10 sources undetected in the Ueda et al.
catalogue were found to have fitted values consistent with background noise,
even in the hard band (see Table~1).  We therefore find that 83\% of our
sources classified as high-redshift type-2 quasars are undetected in the X-ray
image, even down to a flux limit of $\sim$3$\times10^{-18}$ W m$^{-2}$ in the
2-12 keV band (the exact limit varies across the X-ray image).

\section{Discussion} 
 
We have found that 10 out of 12 of our sources classified as type-2 quasars,
are undetected in an X-ray image with an approximate flux limit
$\sim$3$\times10^{-18}$ W m$^{-2}$ (in the 2-12 keV band). For the two
detections, we use the photometric redshifts and the Monte-Carlo models of
\citet {1999MNRAS.309..862W} to estimate the absorbing columns. The estimates
for ID347 and ID401 are log$_{10}(N_{\rm H} / \rm m^{2})=$ 27.50 and 27.25
respectively (although ID347 requires a black body at 10$^{7}$ K in addition
to the WF99 spectrum, to fit an upturn in the soft X-rays). For both these
objects, the intrinsic X-ray to mid-infrared ratio seems to be slightly
lower than the E94 expectation by a factor $\sim$2-5.

%Our two X-ray detected objects are at the high-end of the Compton-thin $N_{\rm
%  H}$ range with mid-IR/X-ray flux ratios comparable to those of lower
%redshift objects \citep {2004ApJS..154..160R, 2004A&A...418..465L} of lower
%intrinsic $N_{\rm H}$ because for obscured objects redshifting boosts the
%observed X-ray to near-IR ratio

For the undetected objects, it is not possible to estimate values of $N_{\rm
  H}$. However, as Figure~1 shows, the sensitivity of the
\textit{XMM-Newton\/} observations and the ``negative K-correction'' in X-rays
mean that our X-ray data are sensitive to heavily obscured Compton-thin quasars, even if,
like ID347 and ID401, at X-ray energies they are intrinsically weaker than our
first expectation from the median SED of E94. Only when log$_{10}(N_{\rm H} /
\rm m^{2})\grtsim$ 28.50, do sources with $L_{\rm bol}\sim10^{40}$ W become
too faint to be detected in the energy bands covering 2-12 keV (rest-frame
$\sim$6-36 keV).

\begin{figure}%[!h] 
\begin{center} 
\psfig{file=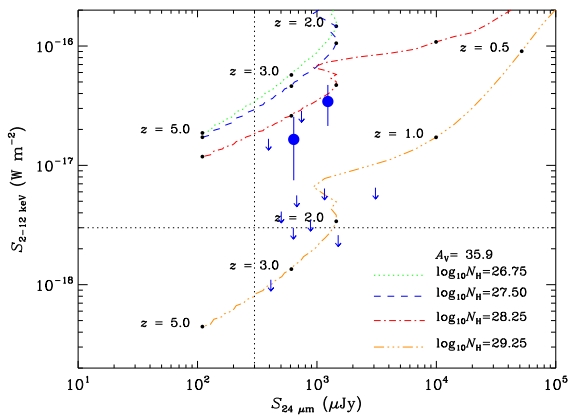,width=8cm,angle=0} 
\caption{\noindent Hard X-ray flux ($S_{2-12 \rm keV}$) versus 24-$\mu$m flux
  density ($S_{24~\mu \rm m}$) as a function of redshift for a model obscured
  quasar. This quasar has been chosen to have log$_{10}(L_{\rm bol}/W)=40.1$
  and $A_{\rm V}=35.9$, which are both the median values from Table~1. The
  quasar has the intrinsic SED of E94, with obscuring dust from the models of
  \citet{1992ApJ...395..130P}. We use the absorbed X-ray spectra of WF99, with
  four different absorbing columns, log$_{10}(N_{\rm H}/ {\rm m}^{-2})=$
  26.75, 27.50, 28.25 and 29.25. This corresponds to a range in gas-to-dust
  ratio of $N_{\rm H}/A_{\rm V}=1.9\times10^{25}$ to $5\times10^{27}$ m$^{-2}$
  (the lower value being that of the Milky Way, the upper value being much
  larger but reasonable, see MS06 and Watanabe et al., 2004). The green and
  blue tracks are obscured but Compton-thin, the red track is mildly
  Compton-thick, while the yellow line is heavily Compton-thick. The black
  dots mark the values at particular redshifts. Changing the $A_{\rm V}$ to
  any other value of Table~1 would move the tracks horizontally only. The 12 
  type-2 quasars in our sample are
  overplotted in blue. The vertical dotted line is the 300 $\mu$Jy flux
  density limit imposed on our sample, the horizontal dotted line is the
  approximate flux limit of the X-ray image, $\sim3\times10^{-18}$ W m$^{-2}$.
  For our two X-ray detections, the independent estimates of log$_{10}(N_{\rm
    H} / \rm m^{2})$ are $\sim27.50$ so at first one expects them to follow
  the blue track. That these two objects lie significantly below the blue
  track is not a cause for serious worry as it assumes a fiducial unobscured
  QSO SED that is subject to population variance.  To make the objects ID347
  and ID401 consistent with the track would mean they have intrinisc
  X-ray to mid-infrared ratios $\sim3\times$ lower than those of the assumed
  SED (E94, see also Figure~1).  This ratio is consistent with the
  $\sim1\sigma$ spread of the population used in E94.
% For our two X-ray detections, the
%  independent estimates of $N_{\rm H}$ are in reasonable agreement with the
%  tracks, although the two sources have X-ray luminosities slightly below the
%  typical expectation (as is also suggested by Figure~1). That these two objects lie significantly
%below the blue track is not a cause for serious worry as it assumes a
%fiducial unobscured QSO SED that is subject to population variance.
%To make the objects ID401 and ID347 consistent with the  track
%would mean they have intrinisc
%mid-infrared to X-ray ratios of x (see also Fig. 1) compared to the
%model value of y, which is within the spread of the population
%as seen in the work of y [nice to be able to say with 1 sigma
%of the spread if this is true!]
  While the two non-detections with high limits are approximately consistent
  with ID347 and ID401 and the Compton-thin tracks, the remaining 8 sources
  are well below the expectations, suggesting they are Compton-thick.  }
\label{fig:xplot} 
\end{center} 
\end{figure}

To get a handle of the possible values of $N_{\rm H}$, in
Figure~\ref{fig:xplot} we show fiducial tracks for a model quasar (see Figure
caption for details). From this figure, the two detected objects in X-rays and
two of the non-detections have values consistent with being heavily absorbed
but Compton-thin, while the other 8 objects undetected in X-rays are likely to
be Compton-thick. Spectroscopic confirmation of the AGN nature
of our sources is now required.

 Inspection of Figure~2 shows most of our objects
populate a region of the $S_{2-12 \rm keV}$ versus $S_{24~\mu \rm m}$ plane
that is lacking in sources in previous work (e.g.  AH06). The reason for this
is that our selection criteria focus in on objects covering a fairly narrow
range in redshift and bolometric luminosity over a fairly large sky area. Most
previous studies (e.g. AH06) cover much larger ranges in z and luminosity with
few objects like ours expected because of small sky area coverage and mid-IR
selection techniques (such as mid-IR-power-law selection) that bias against
the most obscured objects.

%. The
%model quasar has $M_{\rm B}=-25.7$, which from the E94 SED corresponds
%to $L_{\rm bol}$$\sim 7\times10^{39}$ W. Given the relatively narrow
%range in $z$ and $L_{\rm bol}$ of our sample, this track is well
%matched to our candidates and is appropriate (see Figure caption for
%details). %For a sample spanning a larger range in $z$ and $L_{\rm
%  bol}$ (e.g. AH06), it would be meaningless to compare the sample to
%a single track.  
%From this track, all our sources (including our X-ray detections) are
%expected to have log$_{10}(N_{\rm H} / \rm m^{2})\grtsim$ 28.25 and
%therefore would qualify as Compton-thick. This would be consistent
%with our lack of detections. ID347 and ID401 have lower values of
%$N_{\rm H}$, but this is probably due to %intrinsically weaker X-ray
%%emission compared to our fiducial model, as is suggested by Figure~1.
%intrinsically lower X-ray to mid-IR ratios as
%discussed in the caption to Fig. 2. Two of the X-ray non-detections
%have limits consistent with being similar, i.e. log$_{10}(N_{\rm H} / \rm
%m^{2})\sim$27.5-28.25. 

%Two of the non-detections have limits consistent with being heavily
%absorbed but possibly Compton-thin. The remaining 8 objects have much
%lower limits, so if they were heavily obscured but Compton-thin
%quasars, they too would be detected. However, if $N_{\rm H}>10^{28}$
%m$^{2}$, the quasars will be too faint for our observations, even at
%the higher energy bands. Spectroscopic confirmation of the AGN nature
%of our sources is now required.

Following MS05, we model the expected number of type-1 quasars following our
24-$\mu$m and radio flux density criteria, at $z \geq 1.70$, and in 0.8
deg$^{2}$.  We predict 2$\pm0.8$ type-1 quasars following these criteria,
while we have 12 type-2 quasars of which 8 are likely Compton-thick. Assuming
the error in the modelled number of type-1 quasars, and Poisson errors for the
numbers of type-2 quasars, the type-2 to type-1 ratio (in the 68\% confidence
interval) is in the range 3.0-12.9 while the Compton-thick-type-2 to type-1
ratio is in the range 1.8-9.0.

%We have 2 to 4 Compton-thin type-2 quasars
%and possibly between 8 and 10 Compton-thick type-2 quasars. Using
%these numbers, and only the errors in the number of type-1 quasars
%(and in brackets Poisson errors in number of type-1s and type-2s), we
%expect the ratio of Compton-thin type-2 to type-1 quasars to be in the
%range 0.7-3.3 (0.2-10), while the ratio of total type-2 (Compton thin
%and thick) to type-1 quasars is 4.3-10 (2.5-26). It also seems
%$\sim$66-83\% of type-2 quasars are Compton thick, but all these
%numbers are still uncertain.

Although we are clearly suffering from problems due to
small-number-statistics, the implication of our study is that, at high
redshift, Compton-thick quasars may be the dominant sub-population of quasars.
Although such objects would be essentially absent from even moderately hard
X-ray surveys (energy $\sim$2-10 keV), and their contribution to the hard X-ray
background (energy $\geq$10 keV) is diluted by their large distances, they
could clearly represent a vital part of the accretion history of black holes.

Our work to date (MS05, MS06) has included radio as well as mid-IR selection
criteria. There is therefore a residual worry that the properties of the
objects we have studied are influenced in some way by the presence of weak
radio jets, and \citet {2006MNRAS.tmpL.109M} have shown that their radio
properties are consistent with these objects having developed large-scale
FRI-like jets.  In principle, only mid-IR selection criteria are needed to
find high-redshift quasars, although the difficulty lies in filtering out the
starbursts from the quasars.

\section*{Acknowledgments} 
 
%{\scriptsize 
We gladly thank Richard Wilman for access to his X-ray spectra, Filipe Abdalla
for discussions about statistics, Ralf Siebenmorgen for useful comments on the
ULIRG templates and the referee for comments.  SR, DGB and CS thank the UK
PPARC for a Senior Research Fellowship, a Studentship and an Advanced
Fellowship respectively. OA acknowledges the support of the Royal Society.
%}
 
%\bibliographystyle{bibstyle} 
%\bibliography{/home/ams/bibliography/aamnem99,/home/ams/bibliography/references_database} 

\begin{thebibliography}{} 
 
\bibitem[\protect\citeauthoryear{{Alexander}, {Chartas}, {Bauer}, {Brandt}, 
  {Simpson} \& {Vignali}}{{Alexander} et~al.}{2005}]{2005MNRAS.357L..16A} 
{Alexander} D.~M.,  {Chartas} G.,  {Bauer} F.~E.,  {Brandt} W.~N.,  {Simpson} 
  C.,    {Vignali} C.,  2005, \mnras, 357, L16 

\bibitem[\protect\citeauthoryear{{Alonso-Herrero} et~al.,}{{Alonso-Herrero} 
  et~al.}{2006}]{2006ApJ...640..167A} 
{Alonso-Herrero} A.,  et~al., 2006, \apj, 640, 167
 

\bibitem[\protect\citeauthoryear{{Brandt} \& {Hasinger}}{{Brandt} \& 
  {Hasinger}}{2005}]{2005ARA&A..43..827B} 
{Brandt} W.~N.,  {Hasinger} G.,  2005, \araa, 43, 827 
 
\bibitem[\protect\citeauthoryear{{Bruzual} \& {Charlot}}{{Bruzual} \& 
  {Charlot}}{2003}]{2003MNRAS.344.1000B} 
{Bruzual} G.,  {Charlot} S.,  2003, \mnras, 344, 1000 
 
\bibitem[\protect\citeauthoryear{{Cole} et~al.,}{{Cole} 
  et~al.}{2001}]{2001MNRAS.326..255C} 
{Cole} S.,  et~al., 2001, \mnras, 326, 255 
 
\bibitem[\protect\citeauthoryear{{Coleman}, {Wu} \& {Weedman}}{{Coleman} 
  et~al.}{1980}]{1980ApJS...43..393C} 
{Coleman} G.~D.,  {Wu} C.-C.,    {Weedman} D.~W.,  1980, \apjs, 43, 393 
 
\bibitem[\protect\citeauthoryear{{Condon}}{{Condon}}{1992}]{1992ARA&A..30..575% 
C} 
{Condon} J.~J.,  1992, \araa, 30, 575 
 
 
\bibitem[\protect\citeauthoryear{{Donley}, {Rieke}, {Rigby} \& 
  {P{\'e}rez-Gonz{\'a}lez}}{{Donley} et~al.}{2005}]{2005ApJ...634..169D} 
{Donley} J.~L.,  {Rieke} G.~H.,  {Rigby} J.~R.,    {P{\'e}rez-Gonz{\'a}lez} 
  P.~G.,  2005, \apj, 634, 169 
 
\bibitem[\protect\citeauthoryear{{Dwelly} \& {Page}}{{Dwelly} \& 
  {Page}}{2006, DP06}]{2006MNRAS.372.1755D} 
{Dwelly} T.,  {Page} M.~J,  2006, \mnras, 372, 1755 


\bibitem[\protect\citeauthoryear{{Elvis}, {et al.}}{{Elvis} 
  et~al.}{1994}]{1994ApJS...95....1E} 
{Elvis} M.,  et al.,  1994, \apjs, 
  95, 1 
 
\bibitem[\protect\citeauthoryear{{Gilli}, {Comastri} \& {Hasinger}} 
  {{Gilli} et~al.}{2007}]{2006A&AGilli} 
{Gilli} R.,  {Comastri} A.,  {Hasinger} G.,  2007, \aap, 463, 79


\bibitem[\protect\citeauthoryear{Jeffreys}{Jeffreys}{1961}]{Jeffreys:1961} 
Jeffreys H.,  1961, Theory of Probability. 
Oxford Univ. Press 
 

%\bibitem[\protect\citeauthoryear{{Lutz}, {Maiolino}, {Spoon} \& 
%  {Moorwood}}{{Lutz} et~al.}{2004}]{2004A&A...418..465L} 
%{Lutz} D.,  {Maiolino} R.,  {Spoon} H.~W.~W.,    {Moorwood} 
% A.~F.~M.,  2004, \aap, 418, 465
 
 
\bibitem[\protect\citeauthoryear{{Mart{\'{\i}}nez-Sansigre}, {Rawlings}, 
  {Lacy}, {Fadda}, {Marleau}, {Simpson}, {Willott} \& 
  {Jarvis}}{{Mart{\'{\i}}nez-Sansigre} et~al.}{2005}]{2005Natur.436..666M} 
{Mart{\'{\i}}nez-Sansigre} A.,  {Rawlings} S.,  {Lacy} M.,  {Fadda} D., 
  {Marleau} F.~R.,  {Simpson} C.,  {Willott} C.~J.,    {Jarvis} M.~J.,  2005, 
  \nat, 436, 666 
 
\bibitem[\protect\citeauthoryear{{Mart{\'{\i}}nez-Sansigre}, {Rawlings}, 
  {Lacy}, {Fadda}, {Jarvis}, {Marleau}, {Simpson} \& 
  {Willott}}{{Mart{\'{\i}}nez-Sansigre} et~al.}{2006a, hereafter MS06}]{2006MNRAS.tmp..691M} 
{Mart{\'{\i}}nez-Sansigre} A.,  {Rawlings} S.,  {Lacy} M.,  {Fadda} D., 
  {Jarvis} M.~J.,  {Marleau} F.~R.,  {Simpson} C.,    {Willott} C.~J.,  2006a, 
  \mnras,  370, 1479 

\bibitem[\protect\citeauthoryear{{Mart{\'{\i}}nez-Sansigre}, {Rawlings},
    {Garn}, {Green}, {Alexander}, {Kl{\"o}ckner} \&
    {Riley}} {{Mart{\'{\i}}nez-Sansigre} et~al.}{2006b}]{2006MNRAS.tmpL.109M}
  {Mart{\'{\i}}nez-Sansigre} A.,  {Rawlings} S.,  {Garn} T., {Green} D.~A., {Alexander} P.,  {Kl{\"o}ckner} H.-R., {Riley} J.~M., 2006b, 
  \mnras, 373L, 80


\bibitem[\protect\citeauthoryear{{Pei}}{{Pei}}{1992}]{1992ApJ...395..130P} 
{Pei} Y.~C.,  1992, \apj, 395, 130 
 
\bibitem[\protect\citeauthoryear{{Polletta} et~al.,}{{Polletta} 
  et~al.}{2006}]{2006ApJ...642..673P} 
{Polletta} M.~d.~C.,  et~al., 2006, \apj, 642, 673 
 
%\bibitem[\protect\citeauthoryear{{Rigby}  et~al.}{2004}]{2004ApJS..154..160R} 
%{Rigby} J.R.,  et al.,  2004, \apjs, 154, 160 


\bibitem[\protect\citeauthoryear{{Risaliti}, {Maiolino} \& 
  {Salvati}}{{Risaliti} et~al.}{1999}]{1999ApJ...522..157R} 
{Risaliti} G.,  {Maiolino} R.,    {Salvati} M.,  1999, \apj, 522, 157 
 
\bibitem[\protect\citeauthoryear{Siebenmorgen \& Kruegel}{Siebenmorgen \& 
  Kruegel}{2007}]{Siebenmorgen:2007ca} 
Siebenmorgen R.,  Kruegel E.,  2007, \aap, 461,445
 
\bibitem[\protect\citeauthoryear{{Simpson} et~al.,}{{Simpson} 
  et~al.}{2006}]{Simpson:2006if} 
{Simpson} C.,  et~al., 2006, \mnras, 372, 741 


\bibitem[\protect\citeauthoryear{Sivia}{Sivia}{1996}]{Sivia:96} 
Sivia D.~S.,  1996, Data Analysis. A {B}ayesian Tutorial. 
Oxford University Press 
 
\bibitem[\protect\citeauthoryear{{Surace}, {et al.}}{{Surace} 
  et~al.}{2005}]{2005AAS...207.6301S} 
{Surace} J.~A.,  et al. 2005, AAS, 207, 63.01 
 

\bibitem[\protect\citeauthoryear{{Warren}, {et al.}}{{Warren} 
  et~al.}{2007}]{Warren:2006uz} 
{Warren} S.~J.,  et al. 2007, \mnras, 375, 213 

\bibitem[\protect\citeauthoryear{{Watanabe} ,  {et~al.}}{{Watanabe} \& 
  {et~al.}}{2004}]{2004ApJ...610..128W} 
{Watanabe}, C. and {Ohta}, K. and {Akiyama}, M. and {Ueda}, Y.,  2004, \apj,
610, 128


\bibitem[\protect\citeauthoryear{{Wilman} \& {Fabian}}{{Wilman} \& 
  {Fabian}}{1999, WF99}]{1999MNRAS.309..862W} 
{Wilman} R.~J.,  {Fabian} A.~C.,  1999, \mnras, 309, 862 

\end{thebibliography}

\label{lastpage} 
 
\end{document}